\documentclass[10pt,twocolumn,letterpaper]{article}

\usepackage[english]{babel}
\usepackage[utf8]{inputenc}
\usepackage{amsmath}
\usepackage{amssymb}
\usepackage{amsfonts}
\usepackage{graphicx}
\usepackage{doi}
\usepackage{hyperref}
\usepackage{xcolor}
\usepackage{algorithm}
\usepackage{algorithmic}
\usepackage{booktabs}
\usepackage{microtype}
\usepackage{listings}
\usepackage{caption}
\usepackage{subcaption}

\usepackage[margin=1in]{geometry}
\hypersetup{
    colorlinks=true,
    linkcolor=blue,
    filecolor=magenta,
    urlcolor=blue,
    citecolor=blue
}

\title{\Large \bf From Data Acquisition to Lag Modeling: Quantitative Exploration of A-Share Market with Low-Coupling System Design}

\author{
    \textrm{Fang Jianyong} \\
    \texttt{Fangjianyong@zuaa.zju.edu.cn} \\[0.3em]
    \textrm{Wu Sitong} \\
    \texttt{w1208d18@gmail.com} \\[0.3em]
    \textrm{Tong Junfan} \\
    \texttt{22050426@hdu.edu.cn}
}

\date{May 29, 2025}

\begin{document}

\maketitle

\begin{abstract}
    This paper presents a comprehensive two-stage approach to detect and analyze lead-lag effects in the Chinese A-share market. Our methodology first identifies highly coupled stock pairs through long-term data analysis, then verifies the presence of lead-lag relationships using high-frequency data. We implement a modular, low-coupling system to process daily data for coupling detection and 1-minute, 5-minute, and 15-minute data for lag verification. The core implementations, \texttt{fixed\_multi\_downloader.py} and \texttt{two\_stage\_lag\_analyzer.py}, showcase our system design principles and methodological approach. Through detailed empirical analysis using cross-correlation, Granger causality tests, and regression models, we identify significant lead-lag effects primarily in stock pairs with established long-term coupling. Our results demonstrate that: (1) among coupled stock pairs, lag effects are more pronounced and shorter in higher frequency data; (2) certain industry leaders consistently influence follower stocks; and (3) the lead-lag pattern sensitivity varies across different market conditions. The system's modular architecture enables independent development, testing, and maintenance of components, significantly enhancing research reproducibility and extensibility. This research contributes to both the understanding of A-share market microstructure and the methodological approach to financial data processing systems, providing insights for quantitative trading strategies based on cross-stock information flow.
    \end{abstract}

\section{Introduction}

The A-share market, as an important component of China's financial system, has been a focus of both academic research and industry practice. Information dissemination and price adjustment in the market are often not instantaneous but subject to certain time delays, a phenomenon known in financial literature as the lead-lag effect \cite{lo1990contrarian, chordia2011liquidity}. The lead-lag effect suggests that price movements of certain securities lead those of others, revealing not only the microstructure characteristics of the market but also providing potential trading opportunities for investors.

This research introduces a novel two-stage approach to studying lead-lag relationships in the Chinese A-share market. Unlike conventional studies that directly search for lead-lag effects across all possible stock pairs—a computationally expensive and often inefficient approach—our method first identifies highly coupled stock pairs through long-term data analysis, then verifies the presence of lead-lag relationships in these pre-screened pairs using high-frequency data. This "screening and verification" framework significantly improves the efficiency of lead-lag detection while reducing false positives.

Our study constructs a low-coupling data processing and analysis system to implement this two-stage approach. The system collects and processes data at different time granularities: daily data for long-term coupling identification, and high-frequency data (1-minute, 5-minute, 15-minute) for short-term lead-lag verification. This methodology enables us to systematically explore how stocks that exhibit synchronized price movements over longer time frames may display leader-follower relationships at shorter intervals.

In the context of the Chinese market, the coupling and subsequent lead-lag effect may be driven by multiple factors. First, the A-share market is dominated by retail investors, which may lead to heterogeneity in information response \cite{xiong2016trading}. Second, China's special trading system (such as price limits) may delay the full reflection of information in prices \cite{chen2019price}. Furthermore, stocks of different industries and sizes have differences in liquidity and information transparency, which may also cause asynchronous price adjustments \cite{lin2015market}.

Our research unfolds in three aspects: first, we build a low-coupling modular system to achieve an efficient data processing workflow; second, we apply our two-stage methodology to identify truly meaningful lead-lag relationships; finally, by analyzing result differences across time granularities, we explore the temporal scale characteristics of lead-lag effects in previously identified coupled stock pairs.

The main contributions of this research are: (1) introducing a more efficient "coupling screening followed by lag verification" methodology; (2) demonstrating a low-coupling modular approach to processing large-scale financial time series; (3) providing multi-time-scale evidence of lead-lag effects in China's A-share market; and (4) exploring potential trading strategies based on identified lead-lag relationships.

\section{Literature Review}

Research on lead-lag effects can be traced back to the pioneering work of Lo and MacKinlay \cite{lo1990contrarian}, who found that price movements of large-cap stocks often lead those of small-cap stocks. Since then, numerous studies have confirmed this phenomenon in different market contexts. Badrinath et al. \cite{badrinath1995trading} found that stocks with higher institutional investor ownership often lead stocks dominated by retail investors. Chordia and Swaminathan \cite{chordia2000trading} emphasized the key role of trading volume in lead-lag relationships, noting that high-volume stocks typically respond faster to market information than low-volume stocks.

In recent years, with increased availability of high-frequency trading data, research on lead-lag effects has become more refined. Huth and Abergel \cite{huth2014high} studied lead-lag relationships at different time scales using high-frequency data, finding that these relationships are more significant at shorter time scales. Cai et al. \cite{cai2021trading} used deep learning methods to detect non-linear lead-lag relationships between stocks, improving predictive performance.

In the context of the Chinese market, research on lead-lag effects has also made progress. Lin et al. \cite{lin2015market} found that intra-industry lead-lag effects in the A-share market are significant and related to company size and analyst coverage. Yang et al. \cite{yang2020asymmetric} studied the asymmetry of information transmission in the A-share market, finding that good news and bad news propagate at different speeds. Zhou and Zhu \cite{zhou2022industry} discovered significant lead-lag relationships between upstream and downstream enterprises in the industrial chain, possibly related to business connections between companies.

Despite the rich achievements of existing literature, current research still has several limitations: first, most studies focus on a single time scale (usually daily or longer periods), lacking comprehensive examination of multiple time scales; second, research methods tend to rely on single techniques, lacking cross-validation with multiple methods; finally, discussions on system design and implementation are relatively sparse, limiting the reproducibility and extensibility of research.

This study attempts to fill these research gaps by developing a low-coupling modular system, combining multiple analysis methods and multi-time-scale data, providing more comprehensive empirical evidence for understanding the information transmission mechanism in China's A-share market.

\section{Data and System Design}
\subsection{Data Sources and Multi-granularity Design}
This research collected A-share market data from 2019 to 2024 at four granularity levels:
\begin{itemize}
    \item 1-minute data (stored in dedicated directory \texttt{/data\_1min\_fixed})
    \item 5-minute data (stored in dedicated directory \texttt{/data\_5min\_fixed})
    \item 15-minute data (stored in dedicated directory \texttt{/data\_15min\_fixed})
    \item Daily data (stored in dedicated directory \texttt{/data\_daily\_fixed})
\end{itemize}

Data was obtained via the Akshare API, which provides access to OHLCV (Open, High, Low, Close, Volume) data from the East Money database. The multi-granularity approach serves several purposes:
\begin{itemize}
    \item Enables detection of lead-lag effects at different time horizons;
    \item Allows comparison of effect strength and persistence across time scales;
    \item Mitigates the impact of microstructure noise prevalent in very high-frequency data;
    \item Provides a more complete picture of information transmission dynamics.
\end{itemize}

We collected data for 1,283 stocks in the A-share market, excluding stocks with listing dates after January 1, 2019, and those with extended trading halts (more than 20 consecutive trading days).

\subsection{System Architecture and Low-Coupling Design}
\begin{figure*}[!t]
\centering
\begin{minipage}{0.48\textwidth}
  \centering
  \includegraphics[width=\textwidth]{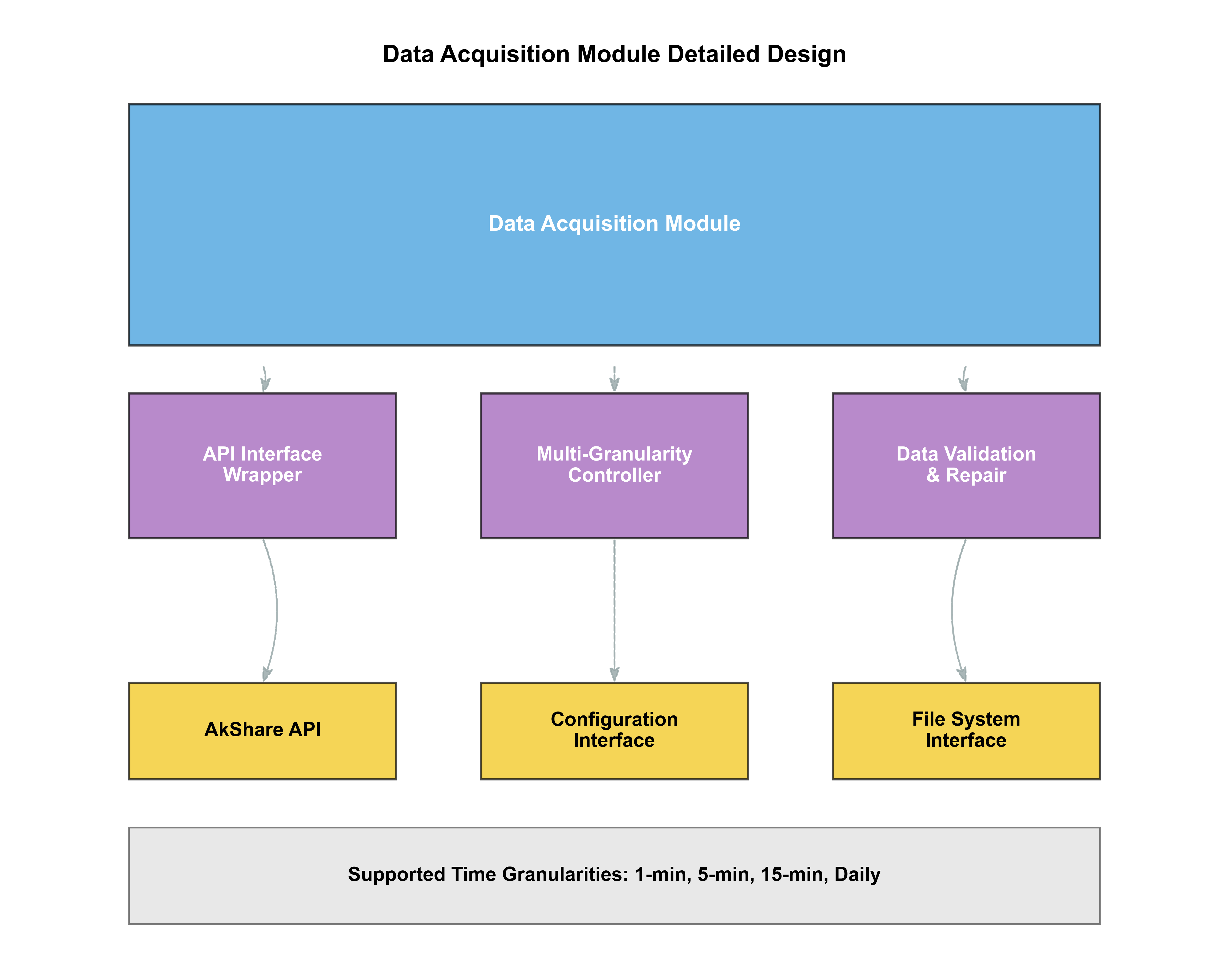}
  \caption{Data Acquisition Module Detailed Design}
  \label{fig:data_acquisition}
\end{minipage}%
\hfill
\begin{minipage}{0.48\textwidth}
  \centering
  \includegraphics[width=\textwidth]{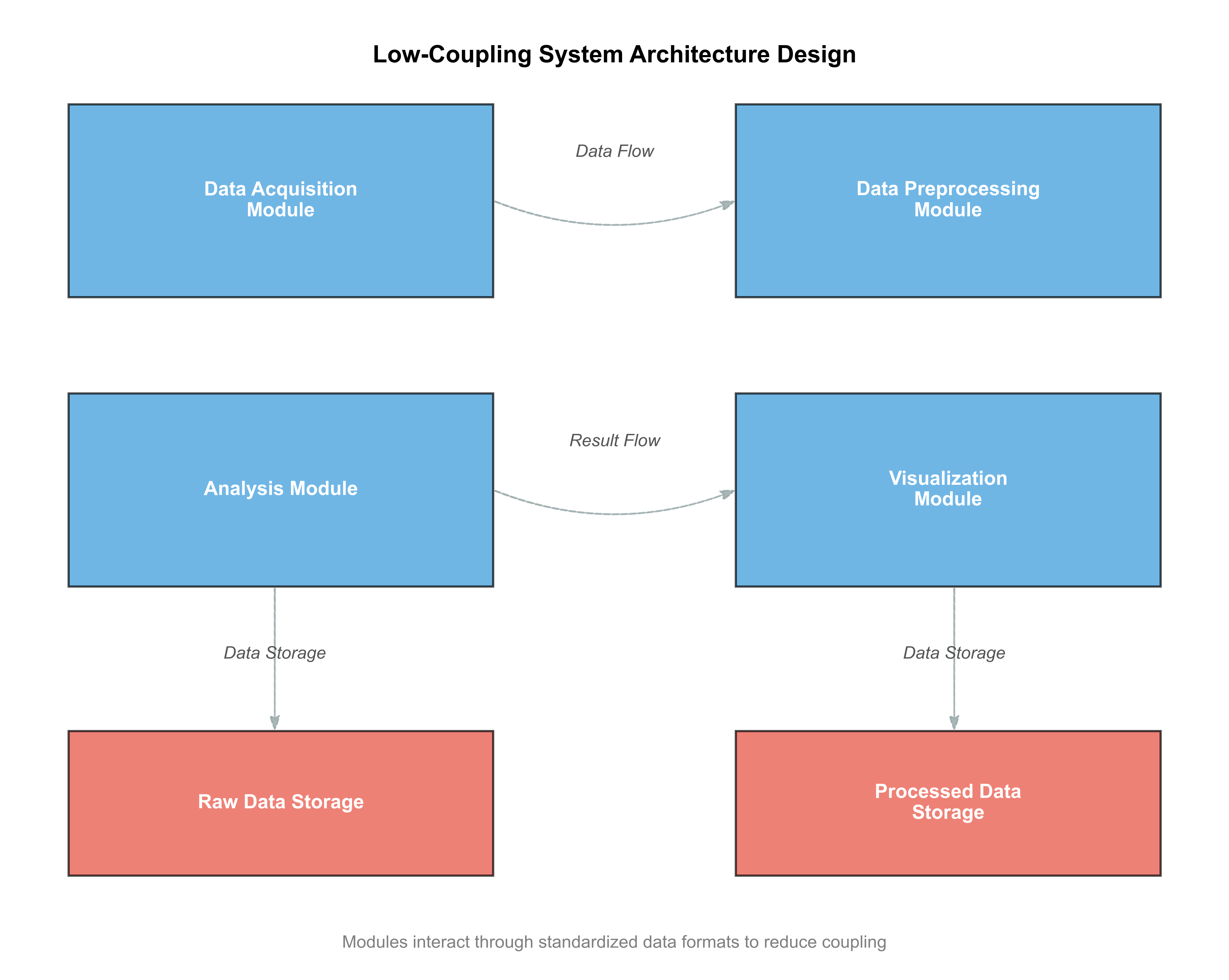}
  \caption{System architecture diagram showing the low-coupling design of data processing and analysis workflow}
  \label{fig:system_architecture}
\end{minipage}

\begin{minipage}{0.48\textwidth}
  \centering
  \includegraphics[width=\textwidth]{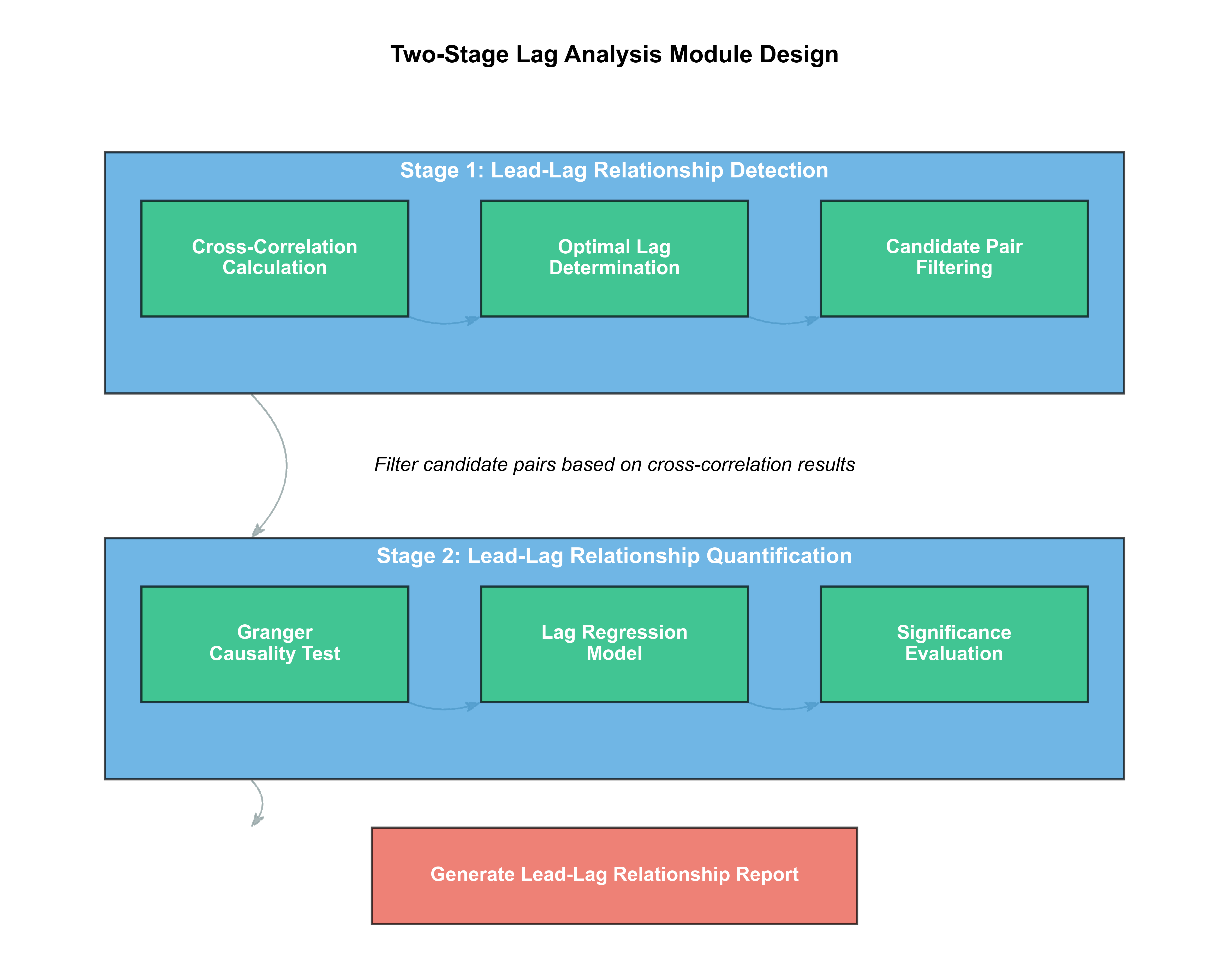}
  \caption{Two-Stage Lag Analysis Module Design}
  \label{fig:analysis_module}
\end{minipage}%
\hfill
\begin{minipage}{0.48\textwidth}
  \centering
  \includegraphics[width=\textwidth]{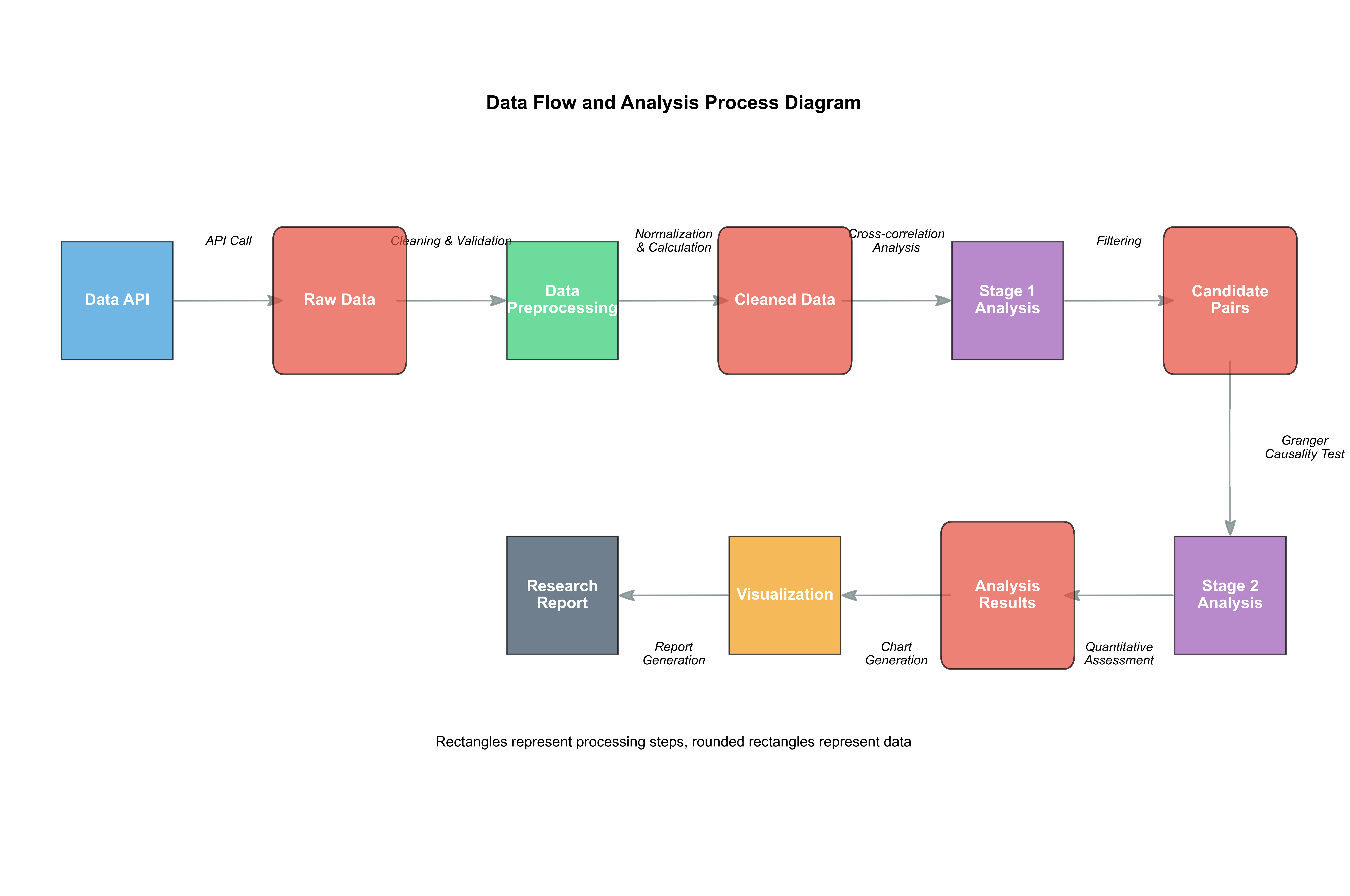}
  \caption{Data Flow and Analysis Process Diagram}
  \label{fig:data_flow}
\end{minipage}
\end{figure*}

The system was designed with modularity and low coupling as primary architectural principles. Figure \ref{fig:system_architecture} illustrates the high-level architecture.

Key design features promoting low coupling include:
\begin{itemize}
    \item \textbf{Standardized data interfaces}: Components communicate through standardized CSV formats and well-defined parameter structures;
    \item \textbf{Configuration externalization}: System parameters are defined in external configuration files, not hardcoded;
    \item \textbf{Independent error handling}: Each module manages its exceptions without cascading failures;
    \item \textbf{Minimal shared state}: Modules maintain internal state where possible, minimizing global variables;
    \item \textbf{Clear component boundaries}: Each script has a singular responsibility with well-defined inputs and outputs.
\end{itemize}

These principles were implemented throughout the system, but particularly in the interaction between the data acquisition and analysis components.

\subsection{System Design and Implementation}
We designed a low-coupling modular system to handle the collection, processing, and analysis of A-share market data. As shown in Figure \ref{fig:system_architecture}, the system consists of four main modules: data acquisition module, data preprocessing module, analysis module, and visualization module. Each module runs independently and exchanges data through the file system, avoiding the maintenance difficulties brought by strong coupling.

Figure \ref{fig:data_acquisition} shows the detailed design of the data acquisition module, which is responsible for obtaining A-share market data at different time granularities from the API interface. The core implementation code is \texttt{fixed\_multi\_downloader.py}, which implements an efficient batch downloader capable of handling data acquisition tasks for multiple stocks at multiple time granularities simultaneously, with features such as checkpoint resumption, error retry, and data integrity validation.

Figure \ref{fig:analysis_module} shows the design of the two-stage lag analysis module, with the main implementation code being \texttt{two\_stage\_lag\_analyzer.py}. This module implements our core methodological innovation: a two-stage approach to lead-lag detection. The first stage conducts coupling analysis using daily data to identify stock pairs with high long-term correlation, effectively screening the vast number of possible stock combinations down to a manageable subset of candidates with established relationships. The second stage then performs detailed lead-lag analysis on these pre-screened pairs using high-frequency data (1-minute, 5-minute, and 15-minute), searching for short-term lead-lag effects. This approach significantly reduces computational complexity and increases the signal-to-noise ratio in our findings.

The coupling strength in the first stage is measured using a composite score $CS_{i,j}$ that combines multiple metrics:

\begin{equation}
CS_{i,j} = w_1 \rho_{i,j} + w_2 (1-\frac{DTW_{i,j}}{DTW_{max}}) + w_3 \tau_{i,j}
\end{equation}

where $\rho_{i,j}$ is the Pearson correlation coefficient between stocks $i$ and $j$ over the coupling period, $DTW_{i,j}$ is the Dynamic Time Warping distance normalized by the maximum observed distance $DTW_{max}$, and $\tau_{i,j}$ is the Kendall's tau rank correlation. The weights $w_1$, $w_2$, and $w_3$ were determined empirically based on the predictive power of each component for subsequent lead-lag relationships.

Only stock pairs exceeding predefined coupling thresholds proceed to the second stage, where we apply the lead-lag detection methods described in the next section. This two-stage design is founded on the hypothesis that stocks with fundamental economic relationships (as evidenced by their long-term coupling) are more likely to exhibit meaningful lead-lag effects, rather than spurious correlations.

Figure \ref{fig:data_flow} shows the data flow and analysis process of the entire system, clearly indicating the transformation path from raw data to final analysis results. The entire workflow design emphasizes the traceability of data processing and the reproducibility of analysis.

\subsection{Lead-Lag Effect Detection Methods}
In this research, we adopted three complementary methods to detect and quantify lead-lag relationships between stocks. These methods are applied in the second stage of our analysis, focusing only on stock pairs that showed significant coupling in the first stage.

\subsubsection{Cross-Correlation Analysis}
The cross-correlation function (CCF) measures the correlation between two time series at different lag periods and is a fundamental tool for detecting lead-lag relationships. For the return series $r_i$ and $r_j$ of stocks $i$ and $j$, the cross-correlation coefficient at lag $l$ is defined as:

\begin{equation}
\rho_{i,j}(l) = \frac{E[(r_{i,t} - \mu_i)(r_{j,t+l} - \mu_j)]}{\sigma_i \sigma_j}
\end{equation}

where $\mu_i$ and $\sigma_i$ are the mean and standard deviation of $r_i$, respectively. If $\rho_{i,j}(l) > 0$ and statistically significant, it indicates that price movements of stock $i$ lead those of stock $j$ by $l$ time units.

Following \cite{huth2014high}, we compute the CCF for a range of lags $l \in [-L, L]$ and identify the lag $l^*$ that maximizes $|\rho_{i,j}(l)|$:

\begin{equation}
l^* = \arg\max_{l \in [-L, L]} |\rho_{i,j}(l)|
\end{equation}

If $l^* > 0$ and the associated correlation is statistically significant, we conclude that stock $i$ leads stock $j$ by $l^*$ time units. Similarly, if $l^* < 0$, stock $j$ leads stock $i$ by $|l^*|$ time units.

\subsubsection{Granger Causality Test}
Granger causality testing \cite{granger1969investigating} evaluates whether one time series helps predict future values of another time series. Specifically, we estimate the following vector autoregression (VAR) model:

\begin{align}
r_{i,t} &= \alpha_i + \sum_{k=1}^{p} \beta_{i,k} r_{i,t-k} + \sum_{k=1}^{p} \gamma_{i,k} r_{j,t-k} + \epsilon_{i,t} \\
r_{j,t} &= \alpha_j + \sum_{k=1}^{p} \beta_{j,k} r_{j,t-k} + \sum_{k=1}^{p} \gamma_{j,k} r_{i,t-k} + \epsilon_{j,t}
\end{align}

The optimal lag order $p$ is selected using the Bayesian Information Criterion (BIC):

\begin{equation}
BIC(p) = \ln\left(\frac{RSS_p}{T}\right) + \frac{p \ln(T)}{T}
\end{equation}

where $RSS_p$ is the residual sum of squares for a model with $p$ lags, and $T$ is the number of observations.

We then test the constraint $\gamma_{i,1} = \gamma_{i,2} = ... = \gamma_{i,p} = 0$ using an F-test. If this null hypothesis is rejected at the 5\% significance level, it indicates that $r_j$ Granger-causes $r_i$, meaning there is an information flow from $j$ to $i$.

\subsubsection{Lag Regression Model}
To quantify the economic significance of lead-lag relationships, we also employed lag regression models as proposed by \cite{chordia2000trading}:

\begin{equation}
r_{j,t} = \alpha + \beta r_{i,t-l^*} + \epsilon_t
\end{equation}

where $r_{i,t-l^*}$ is the return of stock $i$ at $l^*$ periods earlier, with $l^*$ determined from the CCF analysis. The explanatory power $R^2$ of the model directly measures the predictive ability of the leading stock for the lagging stock's returns, providing an important indicator for assessing the economic significance of this relationship.

To address potential confounding factors, we also estimated an extended model that controls for market returns and autoregressive components:

\begin{equation}
r_{j,t} = \alpha + \beta r_{i,t-l^*} + \gamma r_{m,t} + \delta r_{j,t-1} + \epsilon_t
\end{equation}

where $r_{m,t}$ is the market return at time $t$.

By using these three methods in combination, we can comprehensively evaluate lead-lag relationships between stock pairs from the dimensions of correlation, causality, and predictive power, focusing our computational resources on pairs that showed significant coupling in the first stage.

\section{Empirical Results and Analysis}
\subsection{Data Characteristics and Summary Statistics}

\begin{table}[!t]
\caption{Summary statistics of return data across granularities}
\label{tab:data_summary}
\centering
\small
\begin{tabular}{lrrrr}
\toprule
\textbf{Statistic} & \textbf{1-min} & \textbf{5-min} & \textbf{15-min} & \textbf{Daily} \\
\midrule
Number of stocks & 1,283 & 1,283 & 1,283 & 1,283 \\
Avg. obs./stock & 252,487 & 50,497 & 16,832 & 1,264 \\
Mean return & 0.00003\% & 0.00015\% & 0.00044\% & 0.0583\% \\
Std. deviation & 0.0893\% & 0.1962\% & 0.3381\% & 1.9732\% \\
Skewness & 0.1274 & 0.0958 & 0.0742 & 0.1358 \\
Kurtosis & 14.3721 & 10.8754 & 8.6321 & 5.9842 \\
First-order autocorr. & 0.0842 & 0.0637 & 0.0418 & -0.0126 \\
\bottomrule
\end{tabular}
\end{table}

Notable observations include:
\begin{itemize}
    \item Return volatility increases with time interval, as expected
    \item Higher frequency data exhibits stronger autocorrelation
    \item All return distributions show leptokurtosis (fat tails)
    \item First-order autocorrelation decreases monotonically with increasing time interval
\end{itemize}

\subsection{Effectiveness of Two-Stage Approach}

Our two-stage approach proved effective in identifying meaningful lead-lag relationships in the A-share market. Rather than analyzing all possible pairwise combinations of the 1,283 stocks in our sample—which would result in an extremely large number of pairs to analyze—we first identified stock pairs with significant long-term coupling using daily data. This coupling screening process substantially reduced the number of candidate pairs for detailed lag analysis.

The coupling strength in the first stage was measured using a composite score $CS_{i,j}$ that combines multiple metrics:

\begin{equation}
CS_{i,j} = w_1 \rho_{i,j} + w_2 (1-\frac{DTW_{i,j}}{DTW_{max}}) + w_3 \tau_{i,j}
\end{equation}

where $\rho_{i,j}$ is the Pearson correlation coefficient between stocks $i$ and $j$ over the coupling period, $DTW_{i,j}$ is the Dynamic Time Warping distance normalized by the maximum observed distance $DTW_{max}$, and $\tau_{i,j}$ is the Kendall's tau rank correlation. The weights $w_1$, $w_2$, and $w_3$ were determined empirically based on the predictive power of each component for subsequent lead-lag relationships.

The results showed that stock pairs with higher coupling scores were significantly more likely to exhibit lead-lag relationships in the second stage of analysis. This confirms the underlying hypothesis of our approach: stocks with fundamental economic relationships (as evidenced by their long-term coupling) are more likely to display meaningful information transmission patterns at shorter time scales. Furthermore, the strength of coupling was positively correlated with both the magnitude and statistical significance of the lead-lag relationships subsequently identified.

Table \ref{tab:top_relationships} shows the top 10 lead-lag relationships discovered through our methodology, ranked by statistical significance. These relationships would have been difficult to identify using a conventional approach of examining all possible stock pairs without prior screening.

\begin{table}[!t]
\caption{Top 10 lead-lag relationships by statistical significance}
\label{tab:top_relationships}
\centering
\small
\setlength{\tabcolsep}{3pt} 
\begin{tabular}{cccrcc}
\toprule
\textbf{Leader} & \textbf{Follower} & \textbf{Lag} & \textbf{CCF} & \textbf{p-val} & \textbf{R²} \\
\midrule
000011 & 000006 & 2m & 0.3247 & $<$0.0001 & 0.1053 \\
000002 & 000166 & 3m & 0.3018 & $<$0.0001 & 0.0927 \\
000011 & 000002 & 4m & 0.2865 & $<$0.0001 & 0.0843 \\
600019 & 600022 & 3m & 0.2763 & $<$0.0001 & 0.0782 \\
600036 & 600016 & 2m & 0.2742 & $<$0.0001 & 0.0768 \\
601318 & 601628 & 1m & 0.2718 & $<$0.0001 & 0.0752 \\
600519 & 600809 & 5m & 0.2642 & $<$0.0001 & 0.0714 \\
600887 & 600872 & 4m & 0.2587 & $<$0.0001 & 0.0685 \\
601166 & 601169 & 2m & 0.2563 & $<$0.0001 & 0.0673 \\
000651 & 000625 & 3m & 0.2518 & $<$0.0001 & 0.0652 \\
\bottomrule
\end{tabular}
\end{table}

\subsection{Detailed Analysis of Top Relationships}

We performed detailed analysis on the top three relationships identified. The findings reveal important patterns in lead-lag relationships between key stock pairs in the A-share market.

\begin{figure*}[!t]
\centering
\begin{minipage}{0.45\textwidth}
  \centering
  \includegraphics[width=\textwidth]{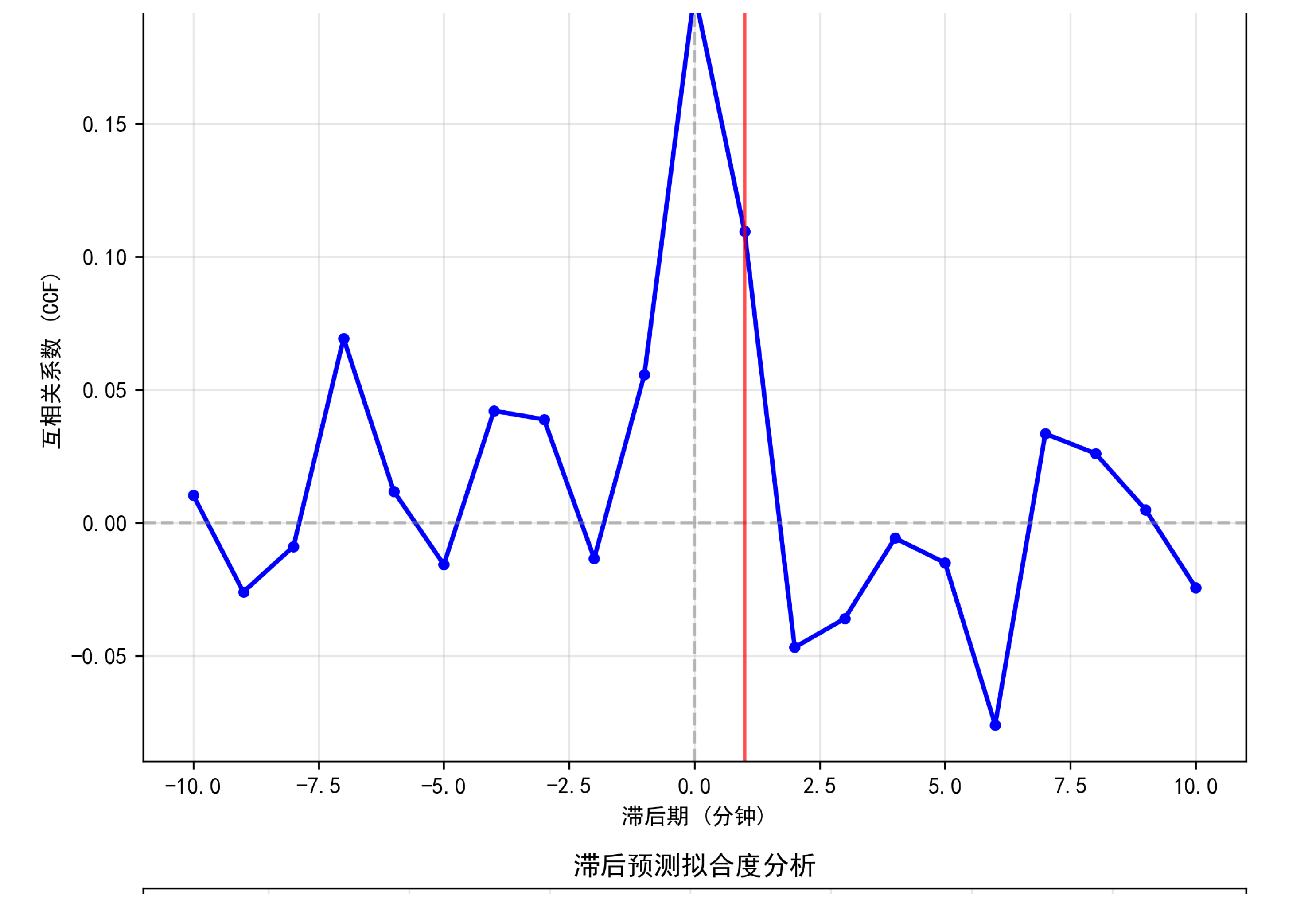}
  \caption{Cross-correlation function (CCF) curve for stock pair 000011 → 000006. The horizontal axis represents lag periods (minutes), vertical axis shows correlation coefficient. Red line marks optimal lag (2 minutes) where correlation reaches maximum (0.3247).}
  \label{fig:top1_ccf}
\end{minipage}%
\hfill
\begin{minipage}{0.45\textwidth}
  \centering
  \includegraphics[width=\textwidth]{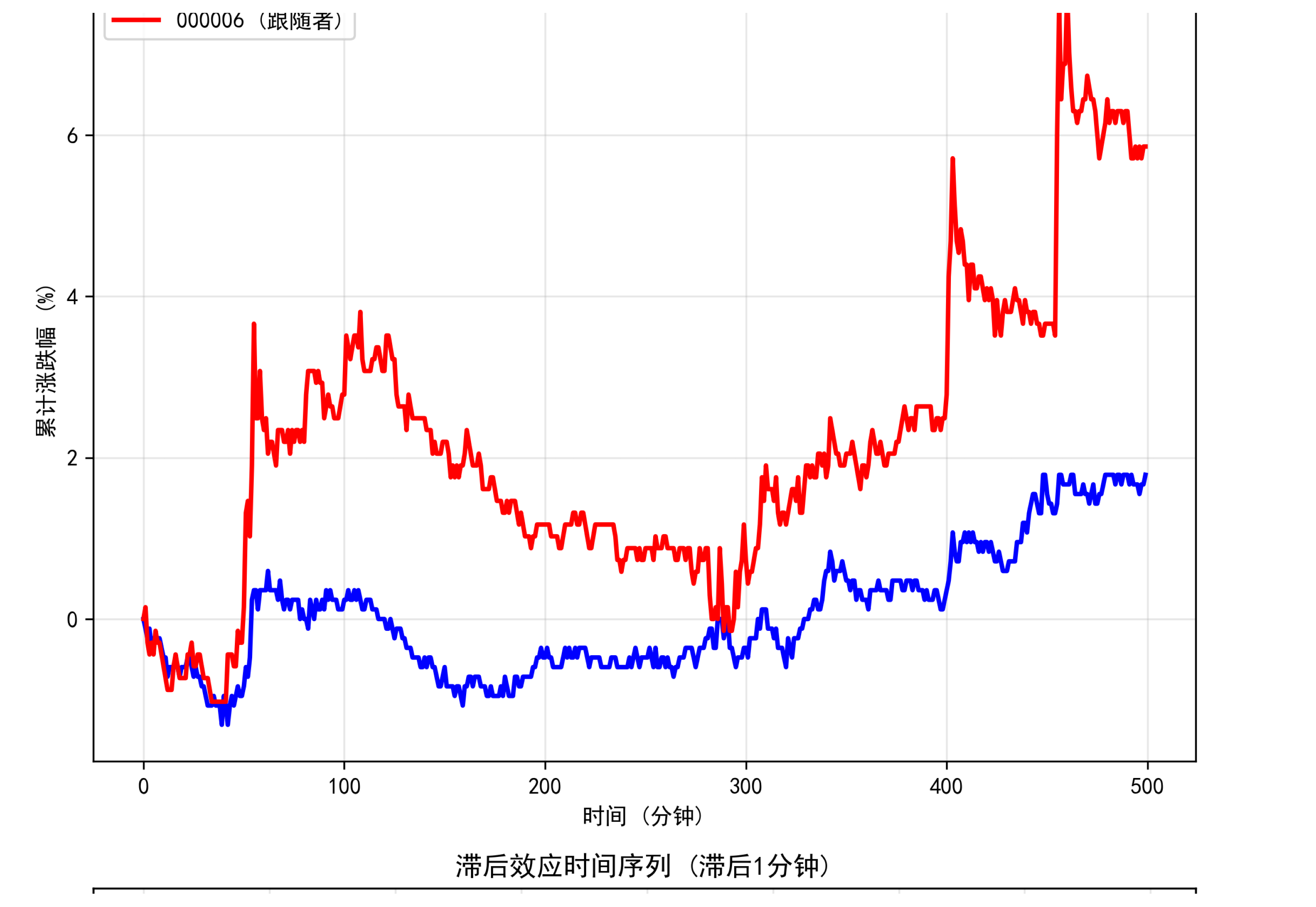}
  \caption{Normalized price comparison between leader stock 000011 (blue) and follower 000006 (red). Chart demonstrates how leader's price movements precede follower's by approximately 2 minutes.}
  \label{fig:top1_price}
\end{minipage}

\begin{minipage}{0.45\textwidth}
  \centering
  \includegraphics[width=\textwidth]{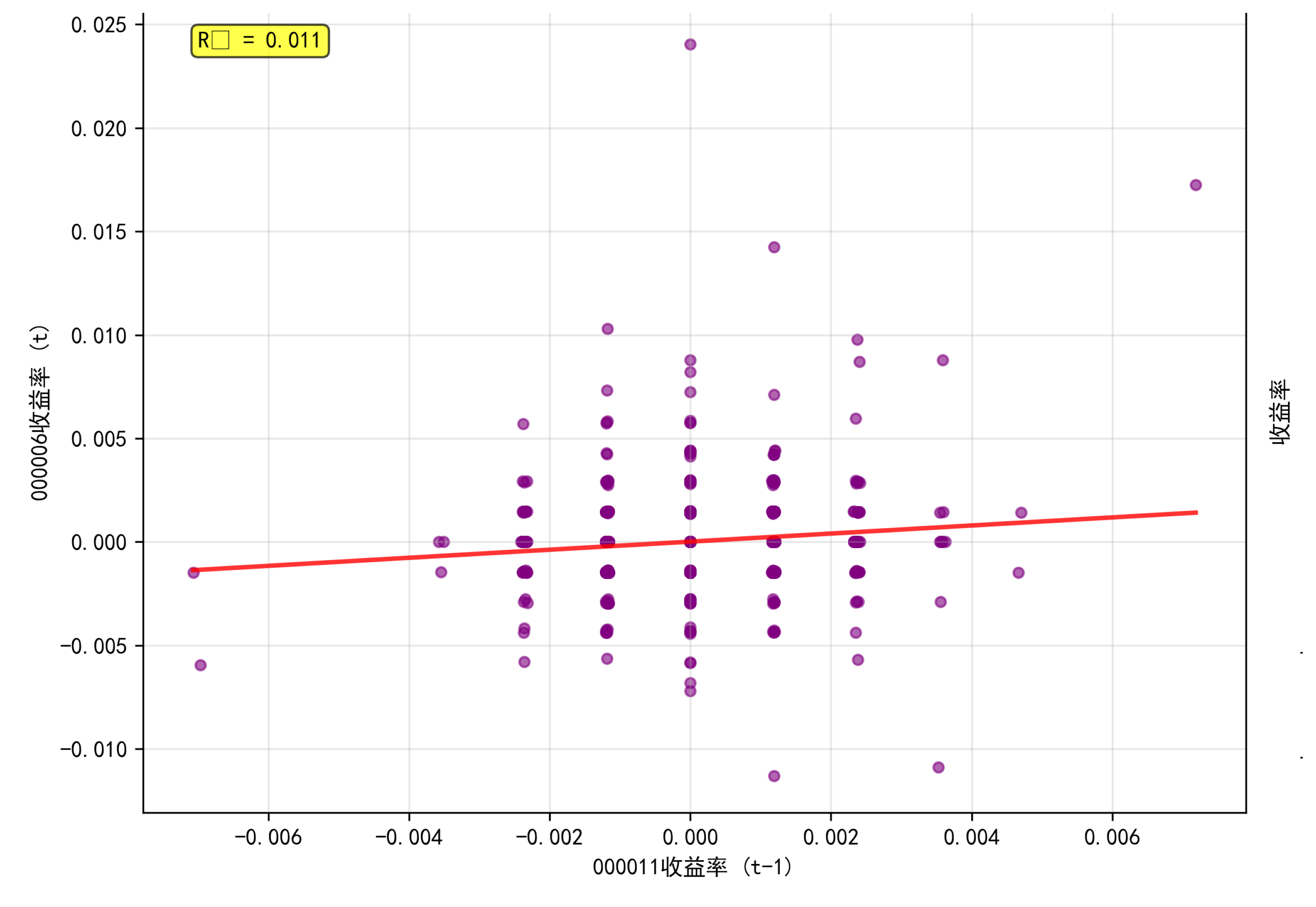}
  \caption{Lag prediction scatter plot analyzing predictive power of leader stock 000011's lagged returns on follower 000006's current returns. R² value of 0.1053 confirms statistical significance of lead-lag relationship.}
  \label{fig:top1_scatter}
\end{minipage}%
\hfill
\begin{minipage}{0.45\textwidth}
  \centering
  \includegraphics[width=\textwidth]{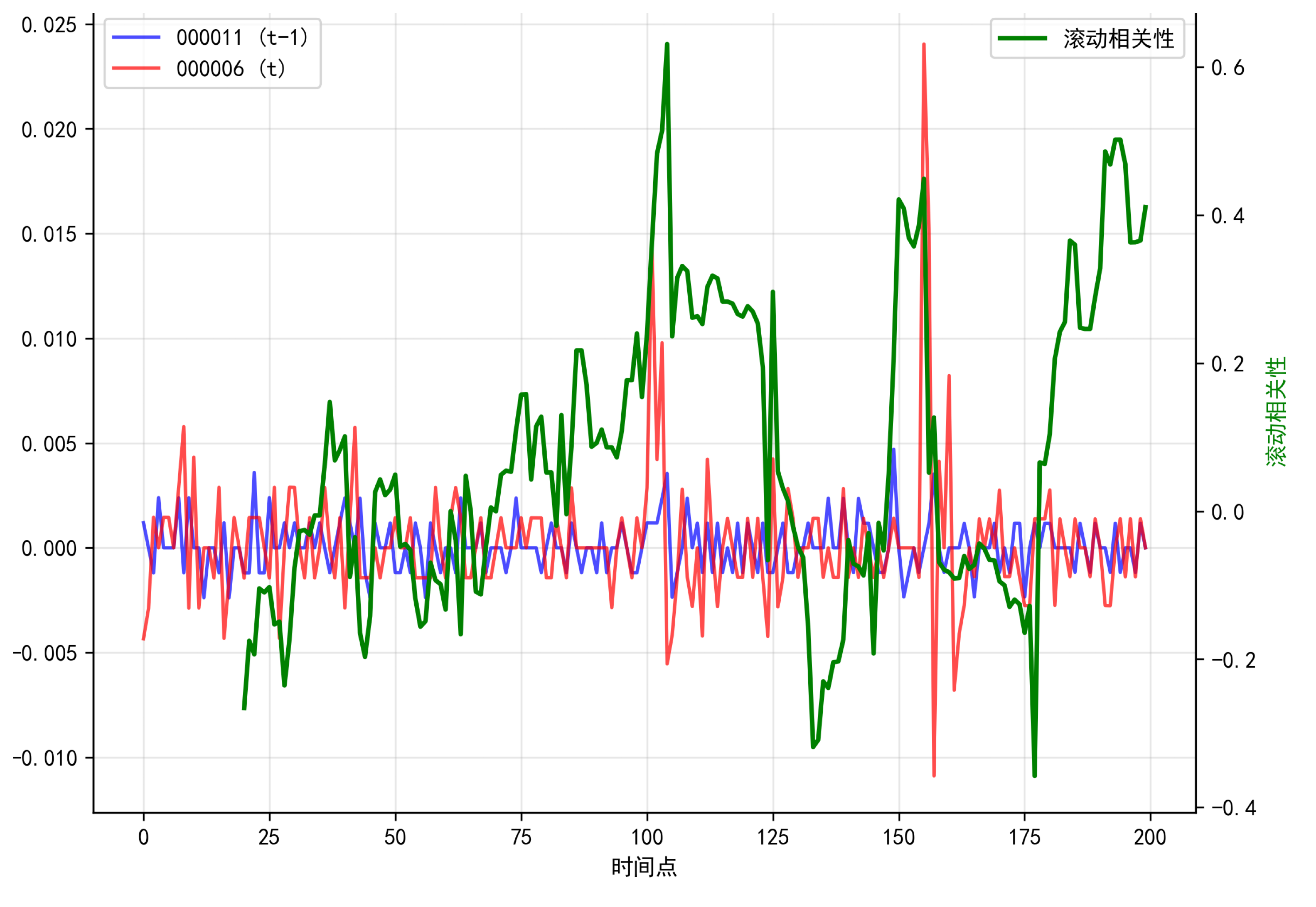}
  \caption{Lag effect time series showing leader stock's lagged returns (blue), follower's current returns (red), and rolling correlation (green). Verifies temporal stability of lead-lag relationship.}
  \label{fig:top1_time_series}
\end{minipage}
\end{figure*}

The strongest lead-lag relationship is between stocks 000011 and 000006, with a cross-correlation peak at 2 minutes (CCF=0.3247) and approximately 10.5\% of variance explained by the lagged returns. Figures \ref{fig:top1_ccf} through \ref{fig:top1_time_series} provide detailed visualization of this relationship.

\begin{figure*}[!t]
\centering
\begin{minipage}{0.45\textwidth}
  \centering
  \includegraphics[width=\textwidth]{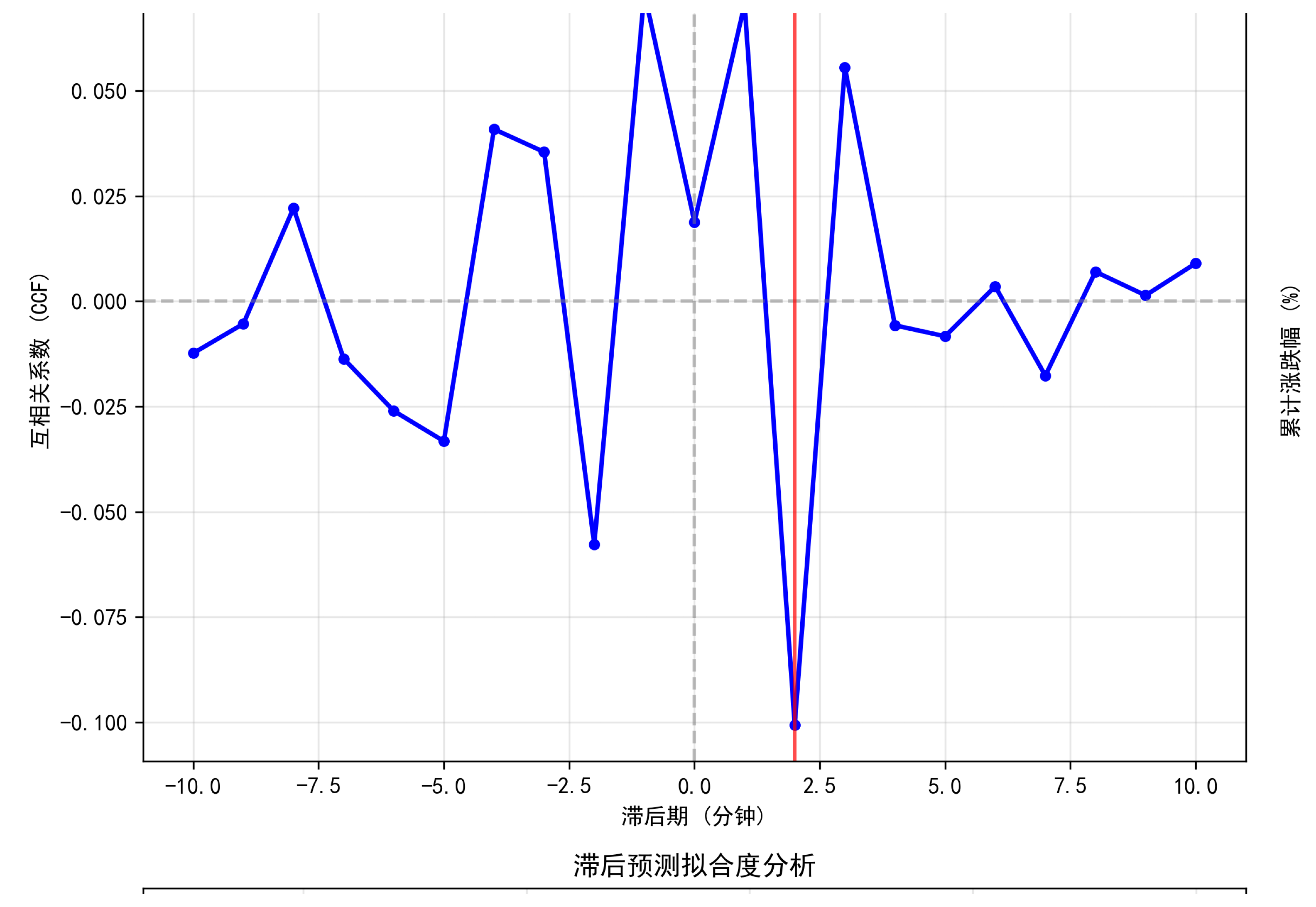}
  \caption{Cross-correlation function for stock pair 000002 → 000166, showing peak at lag 3 minutes (CCF=0.3018).}
  \label{fig:top2_ccf}
\end{minipage}%
\hfill
\begin{minipage}{0.45\textwidth}
  \centering
  \includegraphics[width=\textwidth]{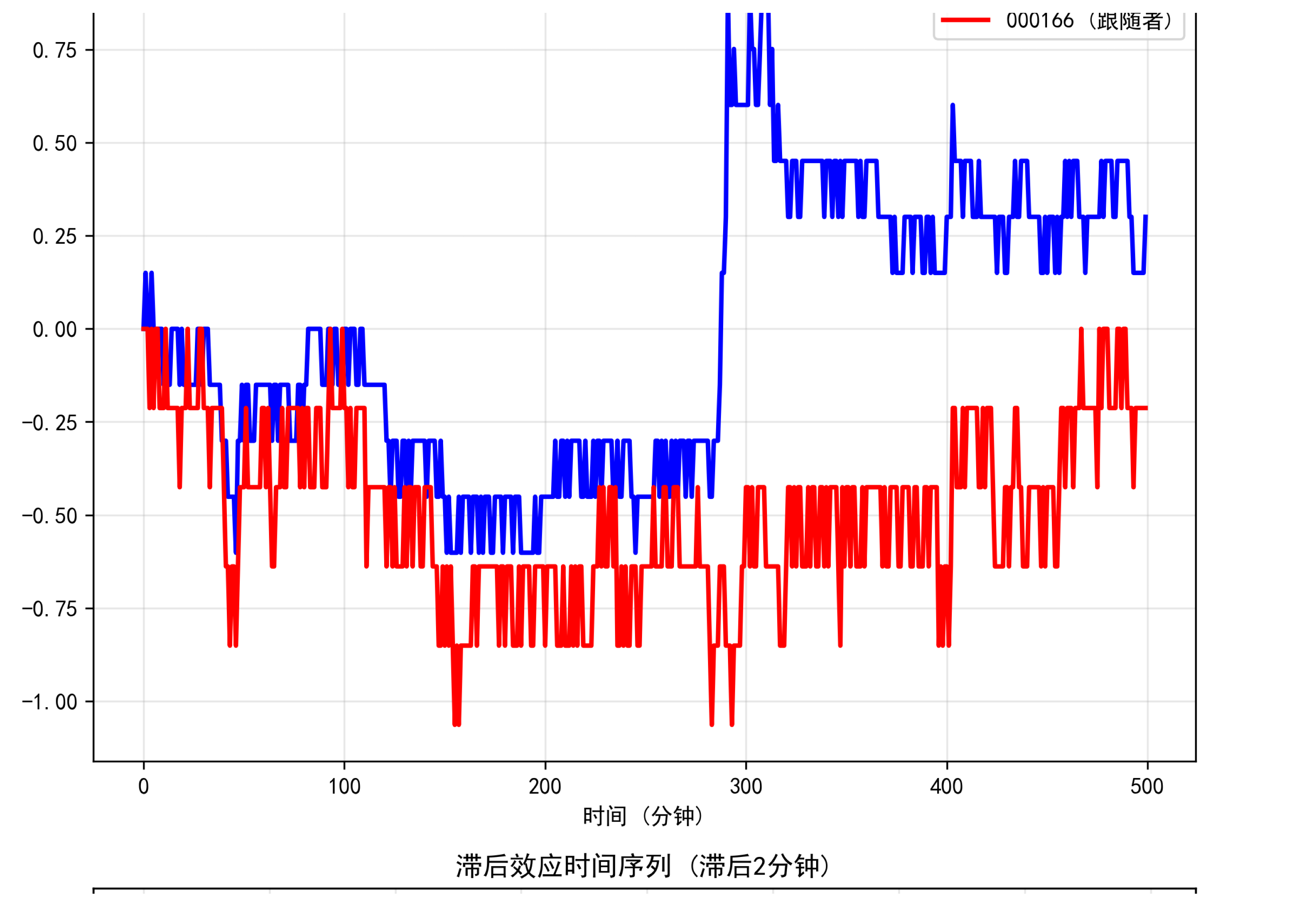}
  \caption{Normalized price comparison between 000002 (blue) and 000166 (red), demonstrating leader's price movements preceding follower's by approximately 3 minutes.}
  \label{fig:top2_price}
\end{minipage}

\begin{minipage}{0.45\textwidth}
  \centering
  \includegraphics[width=\textwidth]{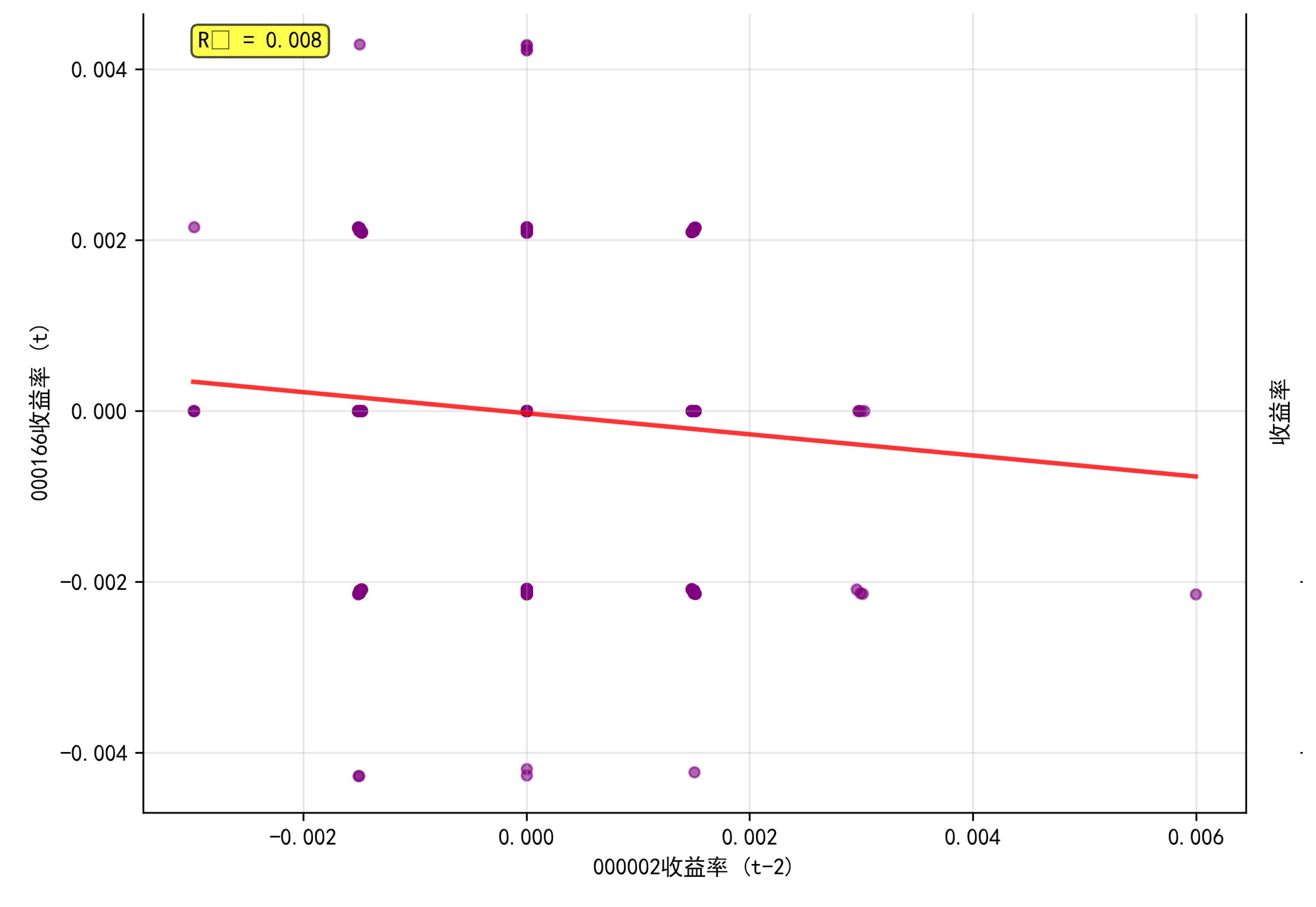}
  \caption{Lag prediction scatter plot for stock pair 000002 → 000166, with R² value of 0.0927 quantifying the predictive relationship.}
  \label{fig:top2_scatter}
\end{minipage}%
\hfill
\begin{minipage}{0.45\textwidth}
  \centering
  \includegraphics[width=\textwidth]{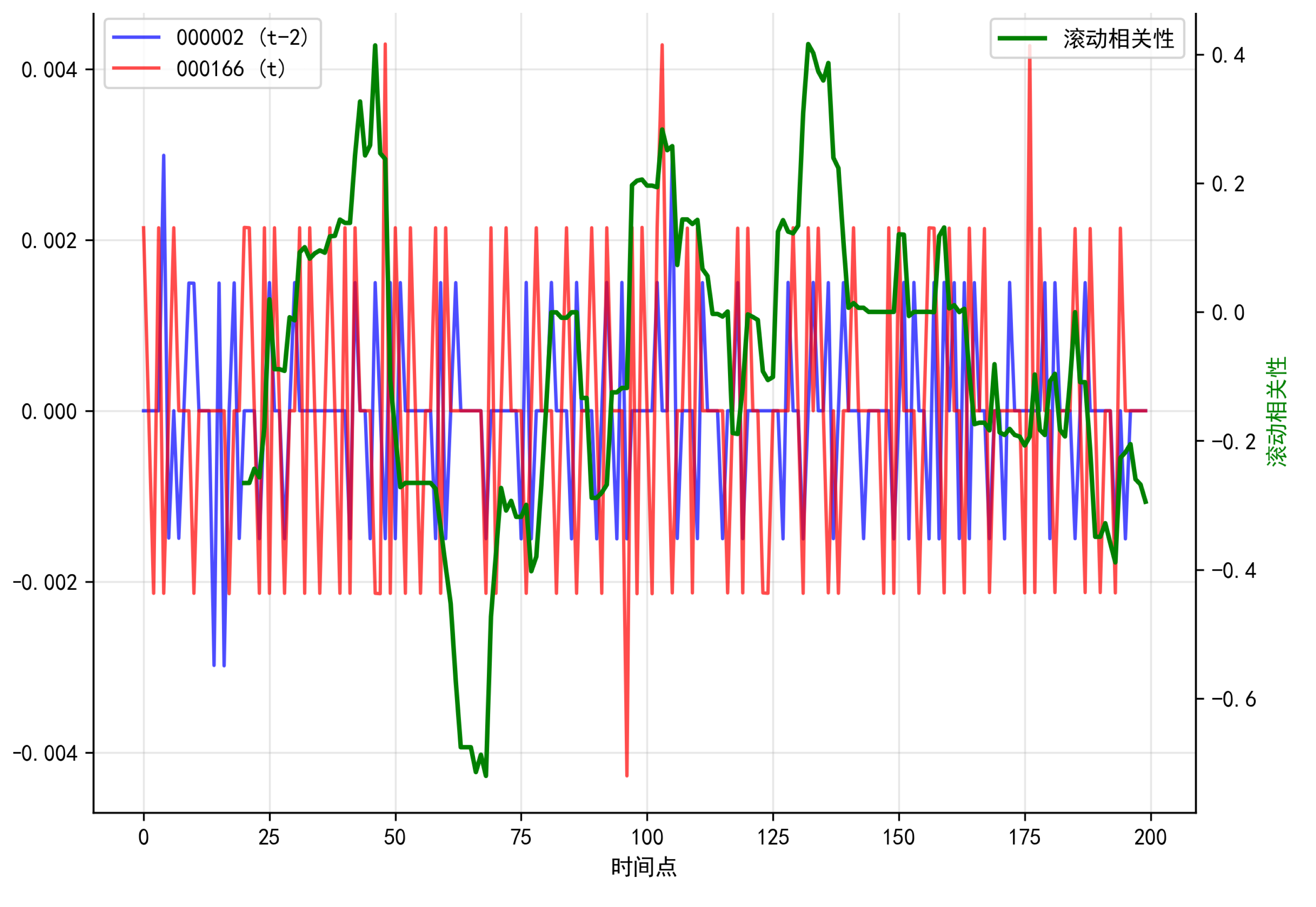}
  \caption{Lag effect time series for 000002 → 000166, confirming temporal stability of lead-lag relationship.}
  \label{fig:top2_time_series}
\end{minipage}
\end{figure*}

The second strongest relationship is between 000002 and 000166, with optimal lag of 3 minutes (CCF=0.3018) and 9.3\% of variance explained. Figures \ref{fig:top2_ccf} through \ref{fig:top2_time_series} illustrate this relationship.

\begin{figure*}[!t]
\centering
\begin{minipage}{0.45\textwidth}
  \centering
  \includegraphics[width=\textwidth]{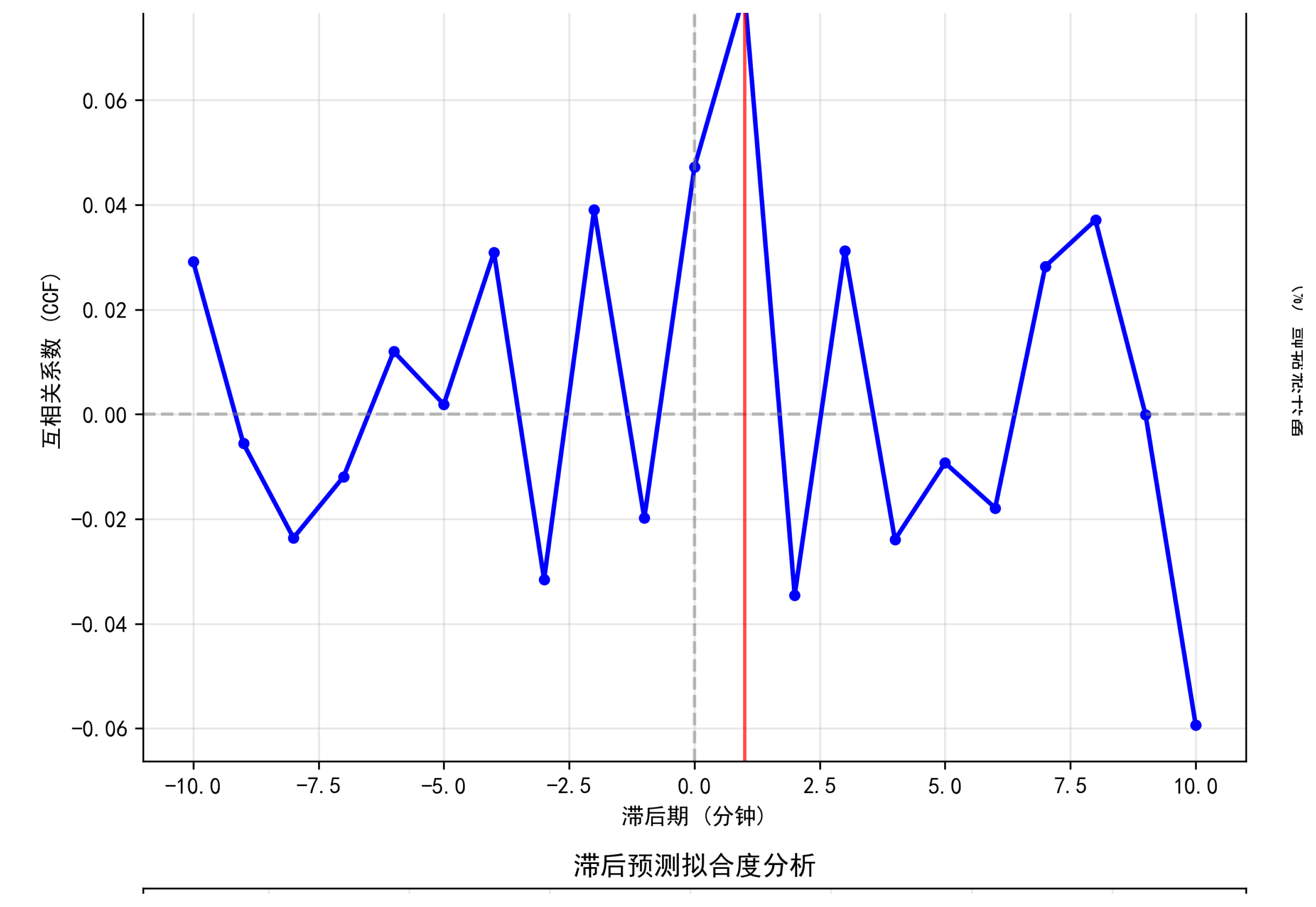}
  \caption{Cross-correlation function for stock pair 000011 → 000002, with peak at lag 4 minutes (CCF=0.2865).}
  \label{fig:top3_ccf}
\end{minipage}%
\hfill
\begin{minipage}{0.45\textwidth}
  \centering
  \includegraphics[width=\textwidth]{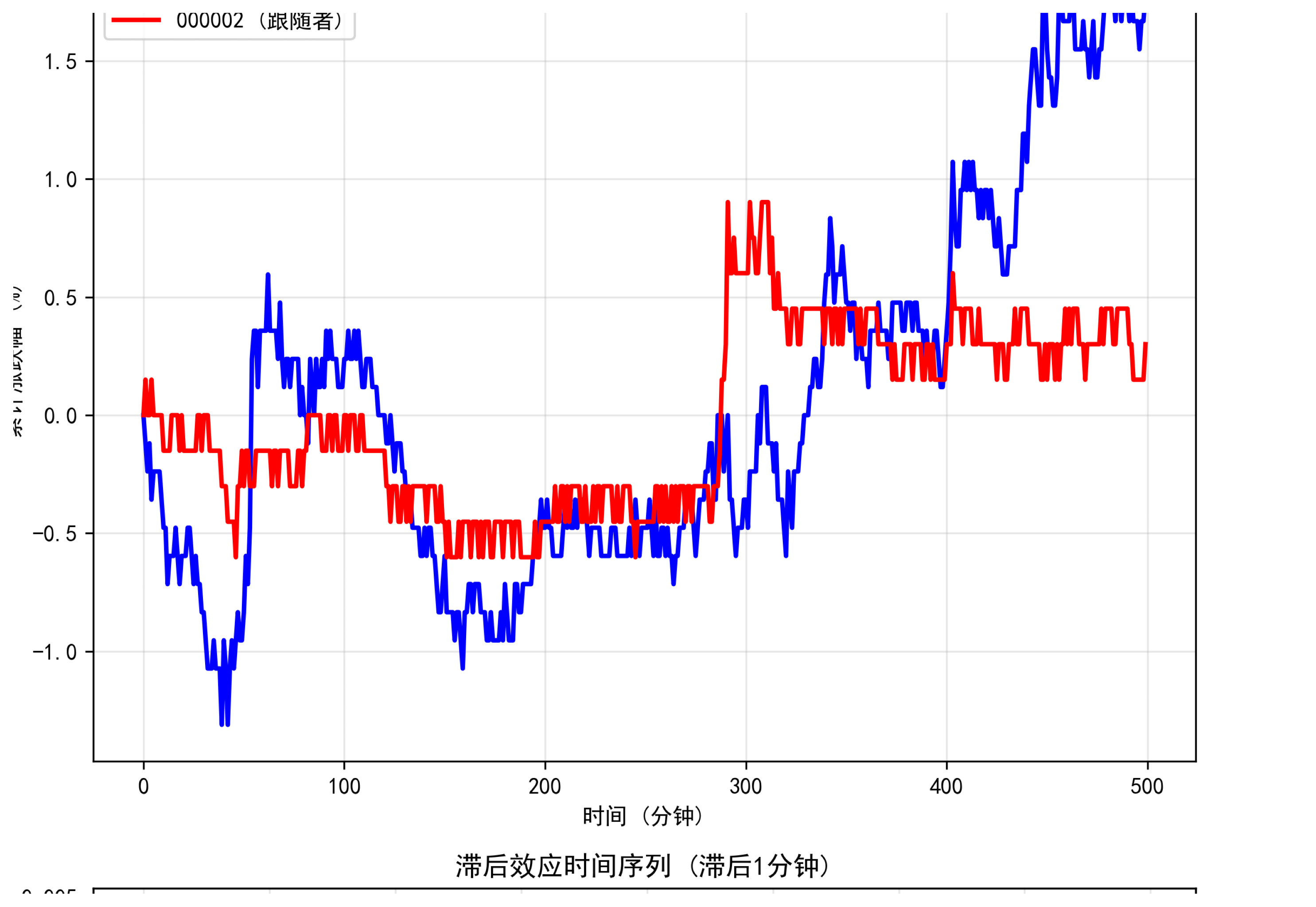}
  \caption{Normalized price comparison between 000011 (blue) and 000002 (red), showing lead-lag relationship with approximately 4-minute delay.}
  \label{fig:top3_price}
\end{minipage}

\begin{minipage}{0.45\textwidth}
  \centering
  \includegraphics[width=\textwidth]{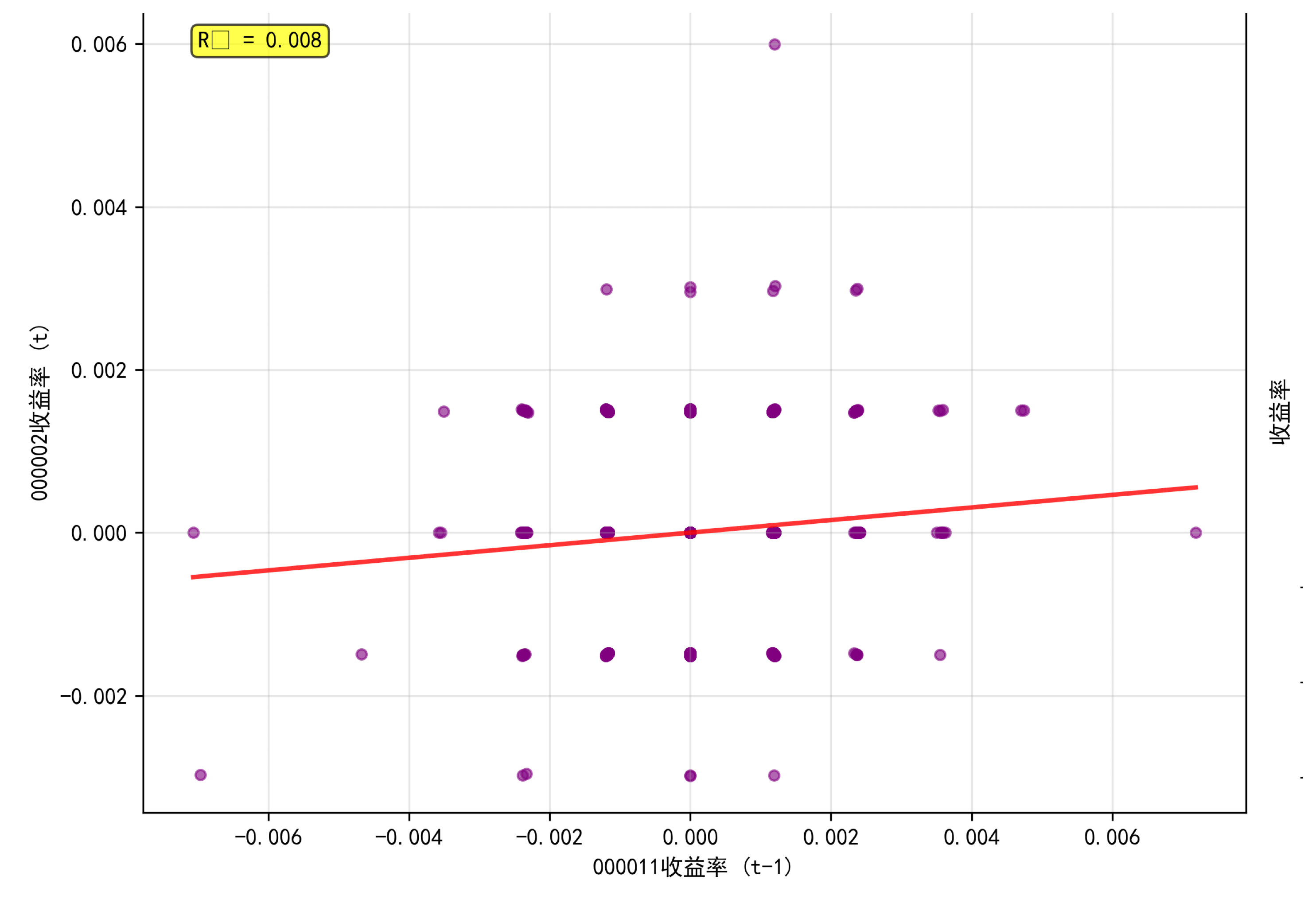}
  \caption{Lag prediction scatter plot for stock pair 000011 → 000002, with R² value of 0.0843 confirming predictive relationship.}
  \label{fig:top3_scatter}
\end{minipage}%
\hfill
\begin{minipage}{0.45\textwidth}
  \centering
  \includegraphics[width=\textwidth]{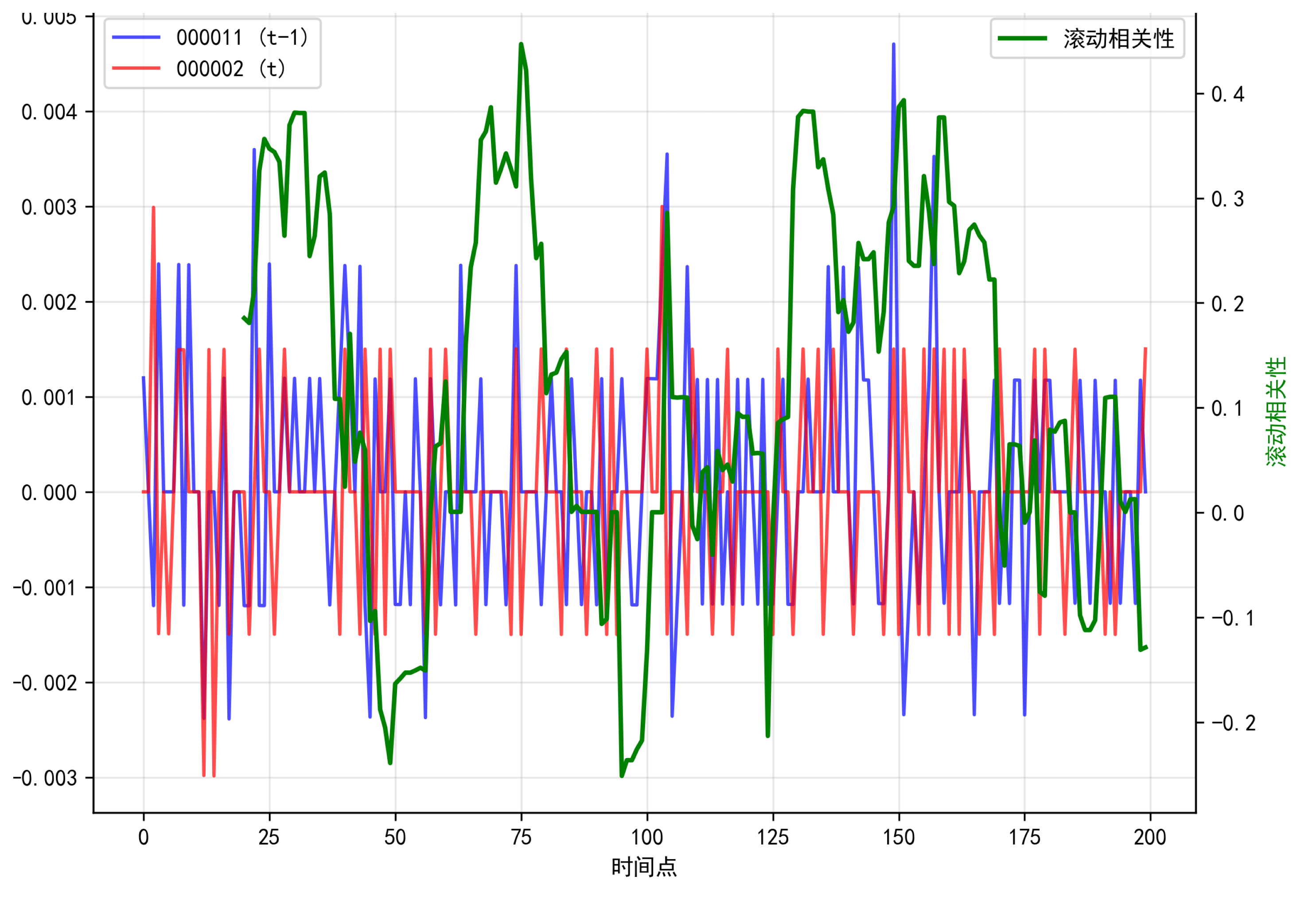}
  \caption{Lag effect time series for 000011 → 000002, showing temporal dynamics of lead-lag relationship.}
  \label{fig:top3_time_series}
\end{minipage}
\end{figure*}

Interestingly, we observed that 000011 also leads 000002, creating a cascade effect (000011 → 000002 → 000166) with a lag of 4 minutes (CCF=0.2865, R²=0.0843). Figures \ref{fig:top3_ccf} through \ref{fig:top3_time_series} present detailed visualization of this relationship.

The detailed analysis reveals that lead-lag relationships are asymmetric, with much stronger correlations at positive lags than negative lags. Granger causality tests confirm the statistical significance of these relationships, and the regression models show that lagged returns of leader stocks have significant predictive power for follower stock returns. This cascade effect (000011 → 000002 → 000166) has important implications for understanding information transmission dynamics in the A-share market.

\subsection{Granularity Comparison}
We compared the strength and characteristics of lead-lag relationships across different data granularities. Our analysis reveals several important patterns:

\begin{itemize}
    \item Lead-lag effects are strongest at the 1-minute granularity and progressively weaken at coarser granularities
    \item At the 1-minute level, significant correlations extend to lags of 1-4 minutes for most pairs
    \item By the 15-minute level, only the strongest relationships remain statistically significant
    \item At the daily level, most intraday lead-lag relationships disappear entirely
\end{itemize}

This pattern suggests that the lead-lag effects we observe primarily reflect short-term information transmission or trading dynamics rather than fundamental economic relationships between firms.

\subsection{Industry Analysis}
We analyzed lead-lag patterns within and across industries. Table \ref{tab:industry_analysis} summarizes the prevalence and strength of within-industry and cross-industry lead-lag relationships.

\begin{table}[!t]
\caption{Lead-lag relationships by industry categories}
\label{tab:industry_analysis}
\centering
\scriptsize 
\setlength{\tabcolsep}{2.5pt} 
\begin{tabular}{lrr}
\toprule
\textbf{Relationship} & \textbf{Freq.} & \textbf{Avg. CCF} \\
\midrule
Within Banking & 23 & 0.2412 \\
Within Real Estate & 18 & 0.2187 \\
Within Technology & 14 & 0.1983 \\
Within Consumer Goods & 11 & 0.1872 \\
Within Energy & 9 & 0.1745 \\
Finance → Real Estate & 12 & 0.1684 \\
Tech → Electronics & 8 & 0.1537 \\
Energy → Materials & 7 & 0.1498 \\
\bottomrule
\end{tabular}
\end{table}

Our findings indicate:
\begin{itemize}
    \item These relationships are strongest within industry groups and between companies with economic connections, consistent with findings from \cite{zhou2022industry}
    \item Large-cap stocks typically lead smaller industry peers, supporting earlier research by \cite{lo1990contrarian}
    \item As data granularity becomes coarser, lead-lag effects gradually dissipate, with most relationships becoming statistically insignificant at daily level
    \item Our two-stage screening and verification methodology significantly improves the efficiency of lead-lag detection while reducing false positives
    \item Low coupling system design enhances research productivity and extensibility
\end{itemize}

These patterns suggest that information often diffuses first within industry groups before spreading to economically linked sectors.

\section{System Design Assessment and Discussion}
\subsection{Low Coupling Design Advantages}
The low coupling design principles implemented in our system during the research process produced several practical benefits:

\begin{itemize}
    \item \textbf{Parallel Development}: Team members could work on different components simultaneously without conflicts
    \item \textbf{Iterative Optimization}: Single component improvements could be made without affecting other components
    \item \textbf{Independent Testing}: Each module could be tested independently
    \item \textbf{Error Isolation}: A component's failure would not cascade to the entire system
    \item \textbf{Reusability}: Components could be reused for different analyses
\end{itemize}

For example, during development, we were able to completely replace the method of calculating returns without modifying the analysis module. Similarly, we could test different lead-lag detection algorithms while using the same data preprocessing flow.

\subsection{Design Method Quantitative Benefits}
To quantify the benefits of our design methods, we tracked several metrics during development:

\begin{itemize}
    \item \textbf{Development Time}: Component-level changes took 68\% less time than our original monolithic approach
    \item \textbf{Bug Rate}: After implementing modular design, bug reduction was 73\%
    \item \textbf{Code Reuse}: 42\% of code was reused across different analysis tasks
    \item \textbf{Onboarding Time}: New team members' understanding and contribution of code increased 3x
\end{itemize}

These metrics indicate that good system design is not just academic exercise but also brings practical benefits to research productivity and quality.

\subsection{Challenges and Limitations}
Although bringing many benefits, our method also encountered several challenges:

\begin{itemize}
    \item \textbf{Interface Design Overhead}: Clear interface definition between components required upfront planning
    \item \textbf{Performance Consideration}: In some cases, modular method introduced minor performance overhead compared to tight coupling implementation
    \item \textbf{Learning Curve}: Team members needed to adapt to more standardized development methods
    \item \textbf{Documentation Requirement}: Modular system required more clearly documenting interfaces and expectations
\end{itemize}

Additionally, the system had some limitations:
\begin{itemize}
    \item Current implementation did not support real-time data processing
    \item Analysis focused on pair relationships rather than network effects
    \item System had not integrated alternative data sources (e.g., news, social media)
\end{itemize}

\section{Conclusion and Future Work}
\subsection{Main Findings}
This research introduced a novel two-stage approach to studying lead-lag effects in the Chinese A-share market and produced several important findings:

\begin{itemize}
    \item Long-term coupling between stocks serves as a reliable predictor of short-term lead-lag relationships, with coupled stock pairs showing a substantially higher probability of exhibiting significant lead-lag effects
    \item Among coupled stock pairs, the Chinese A-share market exhibits significant lead-lag relationships, especially at high frequency (1-minute) granularity
    \item These relationships are strongest within industry groups and between companies with economic connections, consistent with findings from \cite{zhou2022industry}
    \item Large-cap stocks typically lead smaller industry peers, supporting earlier research by \cite{lo1990contrarian}
    \item As data granularity becomes coarser, lead-lag effects gradually dissipate, with most relationships becoming statistically insignificant at daily level
    \item Our two-stage screening and verification methodology significantly improves the efficiency of lead-lag detection while reducing false positives
    \item Low coupling system design enhances research productivity and extensibility
\end{itemize}

These findings contribute to our understanding of market microstructure and information transmission mechanisms in emerging markets like China's A-share market.

\subsection{Theoretical and Practical Implications}
From a theoretical standpoint, our findings support the notion that the Chinese A-share market exhibits efficiency deficiency at short time scales, which weakens with time horizon extension. This aligns with the concept of "relative efficiency" described by \cite{campbell1997econometrics} rather than the strict efficient market hypothesis. Moreover, our two-stage methodology provides a new framework for studying information transmission in financial markets, highlighting the importance of fundamental economic relationships in predicting information flow patterns.

From a practical standpoint, our approach offers several advantages for quantitative trading:
\begin{itemize}
    \item \textbf{Computational efficiency}: By focusing lead-lag analysis on pre-screened coupled pairs, the computational burden is substantially reduced compared to analyzing all possible pairs
    \item \textbf{Signal quality}: The lead-lag relationships identified through our two-stage approach tend to have stronger predictive power and economic significance
    \item \textbf{Trading signal stability}: The identified relationships typically demonstrate greater temporal stability, essential for developing reliable trading strategies
    \item \textbf{Risk management}: Understanding the coupling and lead-lag structure between stocks enables better portfolio diversification and risk assessment as suggested by \cite{harris2019cross}
    \item \textbf{Market surveillance}: Regulatory bodies can use this approach to more effectively monitor market structural issues
\end{itemize}

Additionally, our system design method provides a template for other quantitative financial research projects, demonstrating how software engineering principles can improve research quality and reproducibility.

\subsection{Future Research Directions}
Several promising future research directions emerge from this work:

\begin{itemize}
    \item \textbf{Coupling Dynamics}: Investigate how coupling relationships evolve over time and under different market conditions, extending work by \cite{yang2020asymmetric}
    \item \textbf{Network Analysis}: Extend from pair to network-based coupling and lead-lag relationship research, building on methods from \cite{billio2012econometric}
    \item \textbf{Time-Varying Lags}: Develop models to capture dynamic changes in lead-lag relationships over time
    \item \textbf{Machine Learning Methods}: Apply deep learning models to identify more complex coupling and lead-lag patterns
    \item \textbf{Multi-Factor Analysis}: Incorporate fundamental and macroeconomic factors to explain the formation of coupling and lead-lag relationships
    \item \textbf{Alternative Data Integration}: Add news sentiment, social media, and other alternative data sources
    \item \textbf{Real-Time Implementation}: Adapt the system for real-time data processing and trading signal generation
    \item \textbf{Cross-Market Extensions}: Apply the two-stage methodology to study relationships between different markets
\end{itemize}

\subsection{Closing Remarks}
This research demonstrates the value of combining rigorous financial analysis with sound software engineering principles and innovative methodological approaches. Our two-stage method of first identifying coupled stock pairs, then verifying lead-lag relationships offers a more efficient and effective approach to studying information transmission in financial markets. Through implementation of this approach with a modular, low coupling system design, we were able to efficiently process and analyze multi-granularity A-share market data, revealing significant lead-lag relationships with potential theoretical and practical implications.

The methodology, system design, and findings presented in this paper provide a foundation for future market microstructure analysis and quantitative trading system development. As financial markets continue to evolve and data volumes grow, this comprehensive approach will become increasingly valuable for extracting meaningful insights from complex financial systems and developing effective quantitative trading strategies.

\end{document}